\documentclass[aps,preprint,superscriptaddress,superscriptemail,showpacs,showkeys]{revtex4}
\usepackage[dvipdf]{graphicx}
\usepackage{amssymb}
\usepackage{amsbsy}
\usepackage{latexsym}

\newcommand{\be}{\begin{equation}}
\newcommand{\ee}{\end{equation}}
\newcommand{\ba}{\begin{eqnarray}}
\newcommand{\ea}{\end{eqnarray}}

\begin{document}

\title{A massive graviton in topologically new massive gravity}

\author{Yong-Wan Kim}

\email{ywkim65@gmail.com}

\affiliation {Center for Quantum Spacetime, Sogang University,
Seoul 121-742, Korea}

\author{Yun Soo Myung}

\email{ysmyung@inje.ac.kr}

\affiliation {Institute of Basic Science and School of Computer
Aided Science, Inje University, Gimhae 621-749, Korea}

\author{Young-Jai Park}

\email{yjpark@sogang.ac.kr}

\affiliation {Department of Physics and Department of Global Service Management,\\
Sogang University, Seoul 121-742, Korea}

\begin{abstract}

We investigate the topologically new massive gravity in three
dimensions. It turns out that a single massive mode is propagating
in the flat spacetime, comparing to the conformal Chern-Simons
gravity which has  no physically propagating degrees of freedom.
Also we discuss the realization of the BMS/GCA correspondence.
\end{abstract}



\maketitle

\section{Introduction}

It is well known that the AdS/CFT
correspondence~\cite{Maldacena:1997re} was supported by the
observation that the asymptotic symmetry group of AdS$_3$ spacetime
is two-dimensional conformal symmetry group (two Virasoro algebras)
on the boundary~\cite{Brown:1986nw}. Similarly, the asymptotic
symmetry group of flat spacetime is the infinite dimensional
Bondi-Metzner-Sachs (BMS) group whose dual CFT is described by  the
Galilean conformal algebras (GCA). The latter was called  the
BMS/GCA correspondence~\cite{Bagchi:2010eg}. The centrally extended
BMS (or GCA) algebra is generated by two kinds of  generators $L_n$
and $M_n$:
 \ba\label{GCA}
 &&[L_m, L_n]=(m-n)L_{n+m}+\frac{c_1}{12} (n^3-n) \delta_{n+m,0}, \nonumber \\
 &&[L_m, M_n]=(m-n)M_{n+m}+\frac{c_2}{12} (n^3-n) \delta_{n+m,0}, \nonumber \\
 &&[M_m, M_n]=0.
 \ea

It is very important to establish the BMS/GCA correspondence by
choosing a concrete model. Recently, a holographic correspondence
between a conformal Chern-Simon gravity (CSG) in flat spacetime and
a chiral conformal field theory  was reported
in~\cite{Bagchi:2012yk}. Choosing  the CSG as the flat-spacetime
limit of the topologically massive gravity (TMG) in the scaling
limit of $\mu\to 0$ and $G\to \infty$ while $G\mu$ fixed, the BMS
central charges are determined to be $ c_1=24k(=3/G\mu) $ and
$c_2=0$. This implies that the CSG is dual to a chiral half of a CFT
with $c=24k$. On the other hand, $c_1=0$ and $c_2=3/G$ was predicted
when using the Einstein gravity without taking the scaling
limit~\cite{Barnich:2006av}.

Considering the flat spacetime expressed in terms of outgoing
Eddington-Finkelstein (EF) coordinates,  the linearized equation of
the CSG leads to the third order equation $(Dh)^3=0$.  The  solution
to the first order equation $(Dh^\xi)=0$ is given
by~\cite{Bagchi:2012yk}
\begin{equation}\label{xi-eq} h^\xi_{\mu\nu}=e^{-i(\xi+2)\theta}
r^{-(\xi+2)}(m_1\bigotimes m_1), \end{equation} where $\xi$ is the
eigenvalue of $L_0$ and a Killing vector of
$m_1=ie^{i\theta}(\partial_u-\partial_r-\frac{i}{r}\partial_\theta)$.
Furthermore, one solution to $(Dh)^3=0$ is given by $ h^{\rm
log}_{\mu\nu}=-i(u+r) h^\xi_{\mu\nu}$, while the other is $h^{\rm
log^2}_{\mu\nu}=-\frac{1}{2}(u+r)^2 h^\xi_{\mu\nu}$. These are the
flat-space analogues of log- and log$^2$-solutions on the AdS$_3$
spacetime.

At this stage, we wish to  point out that the solutions
$\{h^\xi,h^{\rm log},h^{\rm log^2}\}$ could not represent any
physical modes propagating on the flat spacetime background
 because the CSG  has no physical degrees
of freedom.  Actually, these all belong to the gauge degrees of
freedom. Hence, it  urges to find a relevant action which has  a
physically massive mode propagating on the Minkowski spacetime.
This might  be found when including  a curvature square
combination $K$, leading to the topologically new massive gravity
(TNMG)~\cite{Andringa:2009yc}.  The TNMG is also obtained from the
generalized massive gravity (GMG) with two different  massive
modes~\cite{Bergshoeff:2009hq,Bergshoeff:2009aq} when turning off
the Einstein-Hilbert term and cosmological constant. If the
Einstein-Hilbert term is omitted, it is called the cosmological
TNMG~\cite{Myung:2012ee}.  It turned out that the linearized TNMG
provides a single spin-2 mode with mass $\frac{m^2}{\mu}$ in the
Minkowski spacetime, which becomes a massless mode of massless NMG
in the limit of $\mu \to
\infty$~\cite{Andringa:2009yc,Dalmazi:2009pm}. Very recently, it
was argued that this reduction (2 $\to$ 1) of local degrees of
freedom is an artefact of  the linearized approximation by using
the Hamiltonian formulation where non-linear effect is not
ignored~\cite{Hohm:2012vh}. We note that the linearized TNMG has a
linearized Weyl (conformal) invariance as the CSG does
show~\cite{Andringa:2009yc}.

In this paper, we explicitly show that a  massive spin-2 mode is
propagating on the flat spacetime when introducing the TNMG.
Furthermore, we observe how the BMS/GCA correspondence is realized
in the TNMG.

\section{TNMG in flat spacetime}

We start with the TNMG action
 \ba
 && I_{\rm TNMG}=I_{\rm CSG}+I_{\rm K},\\
 && I_{\rm CSG} = \frac{1}{2\kappa^2\mu}\int\!\! d^3x\sqrt{-g} \epsilon^{\lambda\mu\nu}
             \Gamma^\rho_{\lambda\sigma}
             \!\!\left(\partial_\mu\Gamma^\sigma_{\rho\nu}+
                   \frac{2}{3}\Gamma^\sigma_{\mu\tau}\Gamma^\tau_{\nu\rho}\right),
                   \label{cgst} \\
 && \label{ikterm}I_{\rm K} = \frac{1}{\kappa^2m^2}\int\!\! d^3x\sqrt{-g}
             \left(R^{\mu\nu}R_{\mu\nu}-\frac{3}{8}R^2\right),
 \ea
where $\kappa^2=16\pi G$, $G$ is the Newton constant, $\mu$ the
Chern-Simons coupling, and $m^2$ a mass parameter. We note that the
GMG action is given by~\cite{Bergshoeff:2009hq,Bergshoeff:2009aq}
\be \label{gmg} I_{\rm GMG}=\frac{1}{16\pi G}\int d^3x\sqrt{-g}
\Big(\sigma R-2 \Lambda_0\Big)+I_{\rm TNMG}, \ee where   the TNMG is
recovered in the limits of $\sigma \to 0$ and $\Lambda_0=-1/\ell^2
\to 0$. The equation of motion of the TNMG action is given by
 \be\label{eom}
 \frac{1}{\mu}C_{\mu\nu}+\frac{1}{2m^2}K_{\mu\nu}=0,
 \ee
where the Cotton tensor $C_{\mu\nu}$  takes the form \be
C_{\mu\nu}=\epsilon_\mu^{~\alpha\beta}\nabla_\alpha
            \left(R_{\beta\nu}-\frac{1}{4}g_{\beta\nu}R\right),
 \ee
and the tensor $K_{\mu\nu}$ is given by
 \ba
 K_{\mu\nu}&=&2\nabla^2R_{\mu\nu}-\frac{1}{2}\nabla_\mu \nabla_\nu R-\frac{1}{2}\nabla^2Rg_{\mu\nu}
          +4R_{\mu\rho\nu\sigma}R^{\rho\sigma} \nonumber\\
        &-&\frac{3}{2} R R_{\mu\nu}-g_{\mu\nu}R_{\rho\sigma}R^{\rho\sigma}
         +\frac{3}{8}{R}^2g_{\mu\nu}.
 \ea

As a solution to Eq. (\ref{eom}), let us choose the Minkowski
spacetime expressed in terms of the outgoing EF coordinates
 \be
 ds^2_{\rm EF}=\bar{g}_{\mu\nu}dx^\mu dx^\nu= -du^2-2drdu+r^2d\theta^2,
 \ee
where $u=t-r$ is a retarded time. Considering the perturbation
$h_{\mu\nu}$ around the EF background $\bar{g}_{\mu\nu}$
 \be
 g_{\mu\nu}=\bar{g}_{\mu\nu}+h_{\mu\nu},
 \ee
the linearized equation of Eq.~(\ref{eom}) takes the form
 \be
 \frac{1}{\mu}\delta C_{\mu\nu}+\frac{1}{2m^2}\delta K_{\mu\nu}=0.
 \ee
Now, we consider  the transverse and traceless conditions  to select
a massive mode propagating on the EF background as
 \be
 \bar{\nabla}^\mu h_{\mu\nu}=0,~~~h^\mu_{~\mu}=0.
 \ee
Then, we have the linearized fourth-order  equation of motion as
 \be
 \epsilon_\mu^{~\alpha\beta}\bar{\nabla}_\alpha\bar{\nabla}^2
     \left(\delta^\rho_\beta+\frac{\mu}{m^2}\epsilon_\beta^{~\sigma\rho}\bar{\nabla}_\sigma\right)h_{\rho\nu}=0,
 \ee
 where the mass of the graviton is identified with $M=m^2/\mu$.
Furthermore, this  equation can be expressed compactly
 \be\label{lin-eom}
\left(D^3D^M h\right)_{\mu\nu}=0
 \ee
by introducing  two mutually commuting operators as
 \be
 D^\beta_\mu=\epsilon_\mu^{~\alpha\beta}\bar{\nabla}_\alpha,
 ~~~(D^M)^\beta_\mu=\delta^\beta_\mu+\frac{\mu}{m^2}\epsilon_\mu^{~\alpha\beta}\bar{\nabla}_\alpha.
 \ee

Now, let us solve the first-order massive equation
 \be\label{meq}
 (D^M h^M)_{\mu\nu}
 =h^M_{\mu\nu}+\frac{\mu}{m^2}\epsilon_\mu^{~\alpha\beta}\bar{\nabla}_\alpha h^M_{\beta\nu}
  \equiv ({\rm EOM})_{(\mu\nu)}=0
 \ee
directly. This will be done  by assuming a proper ansatz
 \be\label{ansatz}
 h^M_{\mu\nu}(u,r,\theta)=f(\theta)G(u,r)
  \left(
  \begin{array}{ccc}
   0 & 0 & 0 \\
    0 & F_{rr}(r) & F_{r\theta}(r) \\
    0 & F_{r\theta}(r) & F_{\theta\theta}(r) \\
  \end{array}
 \right).
 \ee
Then, the traceless  condition of $h^\mu_{~\mu}=0$ takes the form
 \be \label{tlc}
 r^2F_{rr}+F_{\theta\theta}=0,
 \ee
while the transverse conditions $\bar{\nabla}^\mu h_{\mu\nu}=0$
lead to
 \ba
 0 &=& F_{\theta\theta}G f'+
      r f\Big[r GF'_{r\theta}+F_{r\theta}(G+r\partial_rG-r\partial_uG)\Big],\nonumber\\
 0&=& F_{r\theta}Gf'+
       f\Big[\frac{G}{r}(r^3F_{rr}'-F_{\theta\theta})+rF_{rr}(G+r\partial_rG-r\partial_uG)\Big],
 \ea
for $\nu=\theta,~r$, respectively, and for $\nu=u$, it vanishes.
Here the prime (${}^\prime$)  denotes the differentiation with
respect to its argument.

The  nine equations of motion take the following forms:
 \ba
 0&=&({\rm EOM})_{(11)}=({\rm EOM})_{(21)}=({\rm EOM})_{(31)},\nonumber\\
 0&=&\label{eqq1} ({\rm EOM})_{(12)} = -rF_{rr}G f'+f\Big[rGF'_{r\theta}+F_{r\theta}(G+r\partial_rG-r\partial_uG)\Big],\\
 0&=&\label{eqq2}({\rm EOM})_{(13)}
  = rF_{r\theta}G f'+ f\Big[r^2GF_{rr}-rGF'_{\theta\theta}
      +F_{\theta\theta}(G-r\partial_rG+r\partial_uG)\Big], \\
 0&=& \label{eqq3}({\rm EOM})_{(22)} = m^2r^2F_{rr}G f
  + \mu G\Big[-rF_{rr}f'+f(F_{r\theta}+rF'_{r\theta})
      +\mu r fF_{r\theta}\partial_rG)\Big],\\
 0&=&\label{eqq4}({\rm EOM})_{(23)}
  = \mu r fF_{\theta\theta}\partial_rG \! -\! G\Big[\mu rF_{r\theta}f'\!+\!f(\mu r^2F_{rr}\!+\!\mu F_{\theta\theta}
    \!-\!r^2(m^2 r F_{r\theta}\!+\!\mu F'_{\theta\theta}))\Big],\\
 0&=& \label{eqq5}({\rm EOM})_{(32)} = F_{r\theta}G-\frac{\mu r}{m^2} F_{rr}\partial_uG,\\
 0&=&\label{eqq6}({\rm EOM})_{(33)}  = F_{\theta\theta}G-\frac{\mu r}{m^2}
  F_{r\theta}\partial_uG
 \ea
with $(u,r,\theta)=(1,2,3)$. From Eq.~(\ref{eqq5}), one finds the
relation
 \be
  F_{r\theta}=\frac{\mu r}{m^2}\frac{\partial_uG}{G}
  F_{rr}.
 \ee
Also, from Eq.~(\ref{eqq6}), one obtains  the relation
 \be
  F_{\theta\theta}=\frac{\mu r}{m^2}\frac{\partial_uG}{G}F_{r\theta}
  =\frac{\mu^2 r^2}{m^4}\frac{(\partial_uG)^2}{G^2} F_{rr}.
 \ee
Comparing this with the traceless condition (\ref{tlc}), we have
 \be
\Bigg[\frac{\partial_uG}{G}\Bigg]^2=- \Bigg[\frac{m^2}{\mu}\Bigg]^2,
 \ee
which could be solved to give
 \be
 G(u,r)=C_1(r)e^{\pm i \frac{m^2}{\mu}u}.
 \ee
Choosing ``$-$"sign, we obtain
 \be
 G(u,r)=C_1(r)e^{-i\frac{m^2}{\mu}u},~F_{r\theta}=-irF_{rr},~F_{\theta\theta}=-r^2F_{rr}.
 \ee
Using these relations, Eqs.~(\ref{eqq1})--(\ref{eqq4}) reduce  to
a single  equation
 \ba
 0=\mu r f C_1(r)F'_{rr}
 -\Big[i \mu C_1(r)f'-\{(2\mu+im^2r)C_1(r)+\mu rC'_1(r)\}f\Big]F_{rr},
 \ea
which has a solution
 \be
 F_{rr}=\frac{e^{-i\frac{m^2}{\mu}r}r^{\frac{if'}{f}-2}}{C_1(r)}.
 \ee
Again, using this, Eqs.~(\ref{eqq1})--(\ref{eqq4}) become a single
equation for $f(\theta)$
 \be
\Big[f'(\theta)\Big]^2=f(\theta)f''(\theta),
 \ee
whose solution  is given by
 \be\label{theta}
 f(\theta)=e^{C_2\theta}
 \ee
 with an  undermined constant $C_2$.

As a result, we arrive at  a solution
 \be \label{fheq}
 h^M_{\mu\nu}(u,r,\theta)=e^{-i\frac{m^2}{\mu}(u+r)}e^{C_2\theta}r^{iC_2-2}
  \left(
  \begin{array}{ccc}
   0 & 0 & 0 \\
    0 & 1 & -ir \\
    0 & -ir & -r^2 \\
  \end{array}
 \right)
 \ee
 with $u+r=t$.
We note that $C_1(r)$ disappears in Eq.~(\ref{fheq}), suggesting
that one may choose $G(u)$ initially, instead of $G(u,r)$ in
Eq.~(\ref{ansatz}).

Importantly, when solving the first-order massive equation
(\ref{meq}), one could not determine $C_2$.  However, if one
introduces the 2D GCA representations, it could be fixed to be
$C_2=-i\xi$, implying that the integration constant $C_2$ can be
interpreted as the eigenvalue of $L_0$ of the asymptotic BMS algebra
in  Minkowski spacetime.  This may be taken as a hint that the TNMG
provides a possible realization of the BMS algebra in three
dimensions.  Making the choice of  $C_2=-i\xi$, we have the solution
 \be \label{tms}
 h^M_{\mu\nu}(u,r,\theta)=e^{-i\frac{m^2}{\mu}(u+r)}e^{-i\xi\theta}r^{\xi-2}
  \left(
  \begin{array}{ccc}
   0 & 0 & 0 \\
    0 & 1 & -ir \\
    0 & -ir & -r^2 \\
  \end{array}
 \right),
 \ee
which is regarded  as our main result.

 In order to
confirm that $h^M_{\mu\nu}$ satisfies the full equation
(\ref{lin-eom}), we apply the massless operator $D$ $n$-times on
$h^M_{\mu\nu}$ as
 \be \label{newrel}
 (D^n h^M)_{\mu\nu}=\left(-\frac{m^2}{\mu}\right)^n h^M_{\mu\nu}.
 \ee
Using Eq.~(\ref{newrel}), it is easy to check that $h^M_{\mu\nu}$
satisfies the linearized fourth-order  equation~(\ref{lin-eom}) as
 \be
 (D^3D^M h^M)_{\mu\nu}=(D^3h^M)_{\mu\nu}+\frac{\mu}{m^2}(D^4 h^M)_{\mu\nu}=0.
 \ee

\section{BMA/GCA correspondence}

We  need more works to confirm that the TNMG provides a possible
realization of the BMS algebra in three dimensions. In this
direction, we may  show that the BMS central charges are  defined in
the TNMG. We might interpret (\ref{tms}) to be one-parameter
deformation of the CSG representation of the the BMS algebra,
parametrized by $M=m^2/\mu$. This is because $
h^M_{\mu\nu}(u,r,\theta)\sim e^{-iM(u+r)} h^\xi_{\mu\nu}(r,\theta)$
where $ h^\xi_{\mu\nu}(r,\theta)$ is the CSG wave function in
(\ref{xi-eq}). Hence, we have to look for the BMS central charges
$c_1$ and $c_2$, eigenvalue $\xi$ of operator $L_0$  and eigenvalue
$\Delta$ of operator $M_0$.

In order to see what is going on  the BMS/GCA correspondence in the
TNMS, we first consider ``the AdS/CFT correspondence on  the AdS$_3$
and its boundary" within the GMG (\ref{gmg}). Two central charges of
the GMG on the boundary are given by
~\cite{Liu:2009pha,Bergshoeff:2012ev,Kim:2012rz}
 \ba
\label{cc1}
c_L&=&\frac{3\ell}{2G}\Big(\sigma+\frac{1}{2m^2\ell^2}-\frac{1}{\mu\ell}\Big),
\\
\label{cc2}
c_R&=&\frac{3\ell}{2G}\Big(\sigma+\frac{1}{2m^2\ell^2}+\frac{1}{\mu\ell}\Big).
 \ea
Taking two limits of $\sigma \to 0$ and $\ell \to \infty$ to obtain
the TNMG, the corresponding BMS central charges are defined to be
 \ba\label{charge1}
  c_1&=&\lim_{\sigma \to 0,\ell \to \infty}(c_R-c_L)=\frac{3}{G\mu},\\
  \label{charge2}
 ~~c_2&=&\lim_{\sigma \to 0, \ell \to\infty}\frac{c_R+c_L}{\ell}=0.
 \ea
Here we observe the disappearance of $m^2$ in $c_2$ (\ref{charge2}),
which might  not be a good news. In defining $c_{1,2}$, we have used
the convention of  ultra-relativistic limit
in~\cite{Bagchi:2012yk,Bagchi:2012xr,Grumiller:2013at}, which is
opposite to $c_1$ and $c_2$ in the original convention of
non-relativistic limit~\cite{Bagchi:2009pe,Setare:2011jt}.  The
former convention  is better to take the flat-spacetime limit from
the AdS$_3$ spacetime.
 Eqs. (\ref{charge1}) and (\ref{charge2})
show clearly that the BMS central charges are determined by the CSG
(\ref{cgst}) solely, implying that the central extensions are
unaffected by the presence of $I_{\rm K}$-term (\ref{ikterm}). This
explains why we have chosen $C_2=-i\xi$ in deriving the massive wave
solution (\ref{tms}).

Now let us determine which one of the rigidity (weight)  $\xi$ and
scaling dimension $\triangle$ is related to the deformed parameter,
mass $M=m^2/\mu$ of the graviton. Since these are eigenvalues as
shown in
 \be\label{eigeneqs}
 L_0|\triangle,\xi>=\xi|\triangle,\xi>,~~
 M_0|\triangle,\xi>=\triangle |\triangle,\xi>,
 \ee
they  are defined by
 \be \xi=\lim_{\ell \to \infty,\sigma \to 0}(h-\bar{h}),
 ~~\triangle=\lim_{\ell \to \infty,\sigma \to 0}\frac{h+\bar{h}}{\ell}.
 \label{xiad}
 \ee
Note that Eq.~(\ref{eigeneqs}) can be understood  as  acting two
operators $L_0=i\partial_\theta$ and $M_0=i\partial_t$ on the
solution (\ref{tms}), defined in Ref.~\cite{Bagchi:2010eg}. Here two
weights $h$ and $\bar{h}$ are defined as the highest weight
conditions of the GMG on the AdS$_3$: ${\cal
L}_0|\psi_{\mu\nu}>=h|\psi_{\mu\nu}>$ and $\bar{\cal
L}_0|\psi_{\mu\nu}>=\bar{h}|\psi_{\mu\nu}>$. Then, $h$ and $\bar{h}$
are determined to be ~\cite{Grumiller:2010tj}
 \be\label{hh}
 (h,\bar{h})=\Big(\frac{3+\ell m_1}{2}, \frac{-1+\ell
 m_1}{2}\Big),
 \ee
where the mass is given by
 \be\label{m-gmg}
 m_1=\frac{m^2}{2\mu}+\sqrt{\frac{1}{2\ell^2}-\sigma
 m^2+\frac{m^4}{4\mu^2}}.
 \ee
According to the ultra-relativistic
convention~\cite{Bagchi:2012xr,Grumiller:2013at},  the connection
between the GCA and the Virasoro algebras is given by
 \be \label{generators}
 L_n={\cal L}_n-\bar{\cal L}_{-n},
 ~~M_n=\frac{{\cal L}_n+\bar{\cal L}_{-n}}{\ell}.
 \ee
After a computation, one finds that
 \be\label{fvalue}
 \xi=2,~~ \triangle=\frac{m^2}{\mu}.
 \ee
The eigenvalue $\xi=2$ arises because it represents spin-2
excitations. In the limit of $\mu \to \infty$, $ \triangle \to 0$ as
recovering the massless NMG.  Using these, the massive wave solution
(\ref{tms}) respects that of the GMG on the AdS$_3$ as
 \be \label{fheq1}
 \tilde{h}^M_{\mu\nu}(u,r,\theta)=e^{-i\frac{m^2}{\mu}(u+r)- 2i\theta }
  \left(
  \begin{array}{ccc}
   0 & 0 & 0 \\
    0 & 1 & -ir \\
    0 & -ir & -r^2 \\
  \end{array}
 \right).
 \ee

At this stage, we note again that the central charges $c_1$ and
$c_2$ in Eqs.~(\ref{charge1}) and (\ref{charge2}) remain  intact in
compared to the CSG, but the scaling  dimension $\triangle$ was
changed from 0 to  $m^2/\mu$. Thinking that the TNMG is
one-parameter deformation of the CSG representation of the BMS
algebra,  one might expect that their central charges also be
deformed. In order to explore this idea, we observe $c_{L/R}$ in
(\ref{cc1}) and (\ref{cc2}) carefully. Considering the
flat-spacetime limit from the AdS$_3$ spacetime, one possibility is
to consider the case
 \be \label{tcc}
 \tilde{c}_2=\lim_{\sigma \to 0,\ell \to \infty}\ell(c_L+c_R)
            =\frac{3}{2Gm^2},
 \ee
while $c_1$ remains the same as in Eq.~(\ref{charge1}). This also
requires a modification of the generator $\tilde{M}_n$  as
 \be
\tilde{ M}_n=\ell({\cal L}_n+\bar{\cal L}_{-n})
 \ee
 instead of $M_n$ in (\ref{generators}).
In this case, the  flat-spacetime definition of the scaling
dimension is changed to be
 \be \label{tdel}
 \tilde{\triangle}=\lim_{\ell \to \infty,\sigma \to 0}\ell(h+\bar{h}),
 \ee
while the rigidity $\xi$ remains unchanged.  For the TNMG, this lead
to the infinity as
 \be
 \tilde{\triangle}=\lim_{\ell \to \infty,\sigma \to 0}\ell(1+\ell m_1)\to \infty,
 \ee
which cannot be acceptable.  As a result, the  BMS charge
$\tilde{c}_2$ seems to be unphysical, even though it has a finite
value (\ref{tcc})  in the flat-spacetime limit.

\section{GMG solution in the flat-spacetime limt}

Now, it is very important to know  what waveform of the
GMG~\cite{Liu:2009pha} provides  (\ref{fheq1}) in the flat-spacetime
limit. Directly, this task will determine which one between
$(c_2=0,\Delta=m^2/\mu)$ and
$(\tilde{c}_2=\frac{3}{2Gm^2},\tilde{\Delta}=\infty)$ is correct. In
particular, the GMG wave solution for the left-moving massive
graviton  is described by
 \be
 \psi^{\rm L}_{\mu\nu} (\rho, \tau^+, \tau^-) = f(\rho, \tau^+, \tau^-)
 \left(
  \begin{array}{ccc}
   1 & 0 & \frac{2i}{\sinh(2\rho)} \\
   0 & 0 & 0 \\
    \frac{2i}{\sinh(2\rho)} & 0 & -\frac{4}{\sinh^2(2\rho)}  \\
  \end{array}
 \right)
 \ee
 in the light-cone coordinates of  ($\rho,\tau^\pm=\tau\pm\phi$)
for the AdS$_3$ spacetime.  Here the amplitude $f$ is given by \be
f(\rho,\tau^+,\tau^-)=e^{-ih\tau^+-i\bar{h}\tau^-}(\cosh\rho)^{-(h+\bar{h})}\sinh^2\rho,
\ee where two weights $h$ and $\bar{h}$ are already given by
(\ref{hh}).  We note that $ \psi^{\rm L}_{\mu\nu}$ satisfies the
traceless and transverse conditions: $\psi^{{\rm
L}\mu}~_{\mu}=0,\bar{\nabla}_\mu \psi^{{\rm L}\mu\nu}=0$.  As is
suggested in Ref.~\cite{Bagchi:2012yk}, we express  the EF
coordinates in terms of global coordinates
 \be
 u=\ell(\tau-\rho),~~r=\ell\rho,~~\theta=\phi.
 \ee
Then, we have a transformed tensor mode
 \ba
 \psi^{\rm L}_{\mu\nu} (u,r,\theta)=f(u,r,\theta)  \left(
  \begin{array}{ccc}
   1 & 1+\frac{2i}{\sinh(\frac{2r}{\ell})} & \ell \\
   1+\frac{2i}{\sinh(\frac{2r}{\ell})}  & 1+\frac{4i}{\sinh(\frac{2r}{\ell})}-\frac{4}{\sinh^2(\frac{2r}{\ell})}  & \left(1+\frac{2i}{\sinh(\frac{2r}{\ell})}\right)\ell \\
   \ell &  \left(1+\frac{2i}{\sinh(\frac{2r}{\ell})}\right)\ell  & \ell^2 \\
  \end{array}
 \right),\nonumber\\
 \ea
where the transformed amplitude takes the form
 \ba
 f(u,r,\theta)=e^{-i\left(\frac{h+\bar{h}}{\ell}\right)(u+r)-i(h-\bar{h})\theta}
 \left[\cosh\left(\frac{r}{\ell}\right)\right]^{-(h+\bar{h})}\sinh^2\left(\frac{r}{\ell}\right).
 \ea
Thus,  taking the flat-spacetime limit of $\ell\rightarrow\infty$
while keeping $u$ and $r$ finite, and  making use of $\Delta$ in
(\ref{xiad}) (but not $\tilde{\Delta}$ in  (\ref{tdel})), we arrive
at
 \be \label{atof}
 \psi^{\rm L}_{\mu\nu}(u,r,\theta)\simeq e^{-i\frac{m^2}{\mu}(u+r) -2i\theta}
 \left(
  \begin{array}{ccc}
   0 & 0 & 0 \\
   0 & 1 & -ir \\
   0 & -ir & -r^2  \\
  \end{array}
 \right),
 \ee
which is exactly the same form of (\ref{fheq1}). This proves that
the massive wave solution (\ref{fheq1}) represents  a truly massive
graviton mode propagating in the Minkowski spacetime background. In
this case, $c_2=0~(\Delta=m^2/\mu)$ is a correct BMS representation
for the TNMG. Finally, we wish to stress that the TNMS provides
one-parameter ($m^2/\mu$) deformation of the CSG representation of
the the BMS algebra. However, the central charges ($c_1,c_2$) are
not affected by this deformation but the scaling dimension $\Delta$
is changed.

\section{Discussions}
Our work was inspired by   the observation that even though the CSG
has  no local degrees of freedom,  it provides the first evidence
for a holographic correspondence (the BMS/CFT
correspondence)~\cite{Bagchi:2012yk}. Its dual field theory is
considered as a chiral CFT with a central charge of $c=24$.

In order to see what happens for the holographic properties of a
gravitational theory with a local degree of freedom, we have
investigated  the TNMG in the Minkowski spacetime. Solving the
first-order massive equation (\ref{meq}) together with the traceless
and transverse conditions, we have found a massive wave solution
(\ref{tms}).  Concerning the BMS/GCA correspondence in the TNMG, we
have $c_1=\frac{3}{G\mu}$ and $c_2=0$ as in the CSG. This means that
the NMG-term ($I_{\rm K}$) does not contribute to the central charge
of the boundary field theory. Also we  have the same rigidity
$\xi=2$ as in the CSG~\cite{Li:2008dq} where
$(h,\bar{h})=(\frac{3+\ell \mu}{2},\frac{-1+\ell \mu}{2})$, but a
different scaling dimension $\Delta=m^2/\mu$ from $\Delta=\mu$ of
the CSG.  Here, some difference arises in defining $\Delta$: in
Ref.~\cite{Bagchi:2010eg}, $\Delta=0$ for the CSG because they have
taken the scaling limit of $\mu \to 0$. However, in this work, we
did not require the scaling limit of $\mu \to 0, G\to \infty$, but
use  the flat spacetime limit of $\sigma \to 0, \ell \to \infty$ to
get the TNMG.  Importantly, we have obtained the massive graviton
wave solution (\ref{fheq1}) which is recovered from the GMG-wave
solution when taking the flat spacetime limit and using $\xi=2$ and
$\Delta=m^2/\mu$.

We discuss asymptotically flat boundary condition on the wave
solution (\ref{fheq1}). Actually,  there is a  difference between
the CSG and the TNMG because there is a change in the radial
solution between  $h^{\xi}_{\mu\nu}$ (\ref{xi-eq}) and
$\tilde{h}^M_{\mu\nu}$ (\ref{fheq1}): $\tilde{h}^M_{\mu\nu}$ is
regular in the interior, but incompatible with the asymptotically
flat boundary condition (3) in Ref.~\cite{Bagchi:2012yk}. Therefore,
there is a little improvement on the radial boundary condition of a
massive graviton mode.

Consequently, we have shown that the TNMG has a single massive mode
propagating on the flat spacetime, whereas there is no physically
propagating degrees of freedom from the CSG.  This means that the
TNMS provides one-parameter deformation of the CSG representation of
the the BMS algebra, parametrized by $m^2/\mu$. However, their
central charges ($c_1,c_2$) are unaffected by this deformation, but
the scaling dimension $\Delta$ is changed.

\begin{acknowledgments}
We would like to thank D. Grumiller for helpful discussions. This
work was supported by the National Research Foundation of Korea
(NRF) grant funded by the Korea government (MEST) through the
Center for Quantum Spacetime (CQUeST) of Sogang University with
grant number 2005-0049409. Y. S. Myung was also supported by the
National Research Foundation of Korea (NRF) grant funded by the
Korea government (MEST) (No.2012-040499). Y.-J. Park was also
supported by World Class University program funded by the Ministry
of Education, Science and Technology through the National Research
Foundation of Korea(No. R31-20002).
\end{acknowledgments}

 \end{document}